TABLE OF CONTENTS (TOC)

**Self-Assembled Formation of Long, Thin, and Uncoalesced GaN Nanowires on Crystalline TiN Films**

David van Treeck,* Gabriele Calabrese, Jelle J. W. Goertz, Vladimir M. Kaganer, Oliver Brandt, Sergio Fernández-Garrido, and Lutz Geelhaar

Paul-Drude-Institut für Festkörperelektronik, Hausvogteiplatz 5–7, 10117 Berlin, Germany

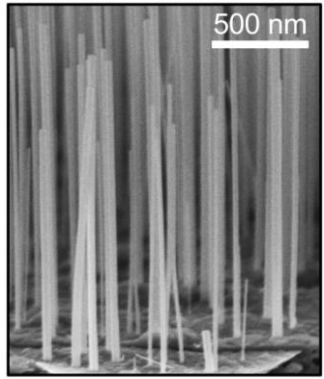
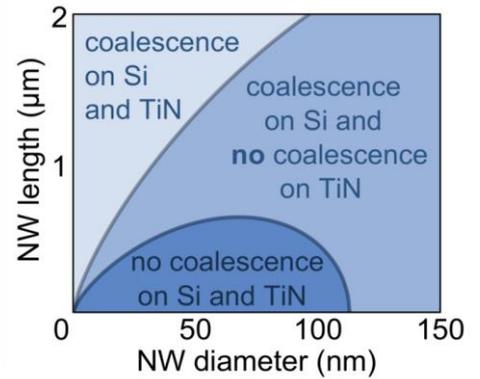

We investigate in detail the self-assembled nucleation and growth of GaN nanowires on TiN. It is demonstrated that this substrate allows the growth of long, thin and uncoalesced GaN nanowires which are very suitable as the basis for the growth of core-shell heterostructures.

\

# Self-Assembled Formation of Long, Thin, and Uncoalesced GaN Nanowires on Crystalline TiN Films

David van Treeck[1](✉), Gabriele Calabrese[1], Jelle J. W. Goertz[1], Vladimir M. Kaganer[1], Oliver Brandt[1], Sergio Fernández-Garrido[1], and Lutz Geelhaar[1]

[1]Paul-Drude-Institut für Festkörperelektronik, Hausvogteiplatz 5–7, 10117 Berlin, Germany

**ABSTRACT**
We investigate in detail the self-assembled nucleation and growth of GaN nanowires by molecular beam epitaxy on crystalline TiN films. We demonstrate that this type of substrate allows the growth of long and thin GaN nanowires that do not suffer from coalescence, which is in contrast to the growth on Si and other substrates. Only beyond a certain nanowire length that depends on the nanowire number density and exceeds here 1.5 µm, coalescence takes place by bundling, i.e. the same process as on Si. By analyzing the nearest neighbor distance distribution, we identify diffusion-induced repulsion of neighboring nanowires as the main mechanism limiting the nanowire number density during nucleation on TiN. Since on Si the final number density is determined by shadowing of the impinging molecular beams by existing nanowires, it is the difference in adatom surface diffusion that enables on TiN the formation of nanowire ensembles with reduced number density. These nanowire ensembles combine properties that make them a promising basis for the growth of core-shell heterostructures.

## Introduction

The self-assembled formation of GaN nanowire (NW) ensembles by plasma-assisted molecular beam epitaxy (PA-MBE) has been achieved for a wide variety of substrates [1–11]. One important advantage of NWs in comparison to planar films is their high crystal quality independent of the lattice mismatch of the substrate [6, 12–16]. This advantage enables, for instance, the integration of GaN-based electronic and optoelectronic devices with technologies based on foreign materials, most notably Si [17–24]. However, on most of the commonly used substrates, such as Si, well developed self-assembled NW ensembles usually exhibit high NW number densities of around $10^{10}$ cm$^{-2}$, leading in turn to a massive coalescence of the NWs [7, 25–30]. The coalescence process can

___________________

Address correspondence to treeck@pdi-berlin.de

result in the formation of boundary dislocations [29, 31–35]. Thus, NW coalescence imposes restrictions on the nanowires' structural perfection, and it is hence desirable to find growth approaches that result in a lower number density and coalescence degree.

For Si and diamond as substrates, it was shown that the NW density can be reduced by high temperature growth, however, as a consequence the NWs are rather short and very inhomogeneous in length [36, 37]. Moreover, further experiments on Si showed that the growth of *long* NWs with *uniform length* at such elevated substrate temperatures again results in high NW densities, similar to growth under conventional conditions [38]. This result is supported by recent findings that the final NW number density on Si is only limited by the shadowing of the impinging fluxes by already existing NWs [30]. An alternative substrate which allows the growth of GaN NW ensembles with lower NW densities is an AlN buffer layer on Si [4, 39–41]. In particular, Brubaker et al. [42] demonstrated the growth of long and thin NWs with densities in the $10^8$ cm$^{-2}$ range on a 40-nm-thick AlN buffer layer. However, on AlN in addition to the NWs a rather thick parasitic GaN layer forms on the substrate [43]. Furthermore, the AlN buffer layer enabling such low NW densities is usually several tens of nm thick and insulating, which is a drawback regarding the use of the NW ensembles in optoelectronic applications. Highly promising, however, seems to be the self-assembled NW growth directly on TiN which has resulted in low number densities [9, 10]. As a metal, TiN has an excellent thermal and electrical conductivity and a high optical reflectivity up to the UV range, which is very beneficial for applications like LEDs based on NW ensembles [44–46]. In a recent investigation, we showed the formation of thin and uncoalesced NWs on crystalline TiN with a NW number density as low as $7\times10^8$ cm$^{-2}$ [9]. At the same time, these NWs were still rather short and varied in length.

In the present study, we demonstrate the self-assembled growth of uncoalesced GaN NWs on TiN with diameters smaller than 50 nm and a homogeneous length of more than 1 µm. Furthermore, we investigate the coalescence behavior in such NW ensembles with different average lengths. We find that coalescence does take place for lengths of a few µm and is described well by the bundling model we recently developed for GaN NW growth on Si [30]. Since according to this model the onset of coalescence is directly related to the NW spacing, we analyze the NW nucleation in an attempt to explain the low number densities on TiN. We conclude that in contrast to nucleation on Si, on TiN the formation of NWs is governed by diffusional repulsion. Besides the suppression of coalescence, the NW ensembles presented here exhibit a combination of number density, length, and diameter that makes them very suitable as the basis for the growth of core-shell heterostructures.

## Experimental Section

The GaN NW ensembles we investigated in this study were grown by means of self-assembly processes with PA-MBE on a Ti film sputtered on a Al$_2$O$_3$ (0001) substrate. The impinging Ga and N fluxes, $\Phi_{Ga}$ and $\Phi_N$, respectively, were calibrated in equivalent growth rate units of planar GaN layers as described elsewhere [47]. We studied three different sets of samples grown under N-rich conditions without rotation: (I) three NW ensembles A, B, and C grown at different substrate temperatures $T_{sub}$ of 750 °C, 725 °C, and 700 °C, respectively, for 4 h on a 2 µm thick Ti film with $\Phi_{Ga}$ = 4.5 nm/min and $\Phi_N$ = 7.8 nm/min,

(II) two NW ensembles D and E grown for 45 min and 6 h, respectively, at 710 °C on a 1.3 μm thick Ti film with $\Phi_{Ga}$ = 4.5 nm/min and $\Phi_N$ = 5.6 nm/min, and (III) one NW ensemble F grown for 7 h at 710 °C on a 1.3 μm thick Ti film, with $\Phi_{Ga}$ = 4.5 nm/min and $\Phi_N$ = 11.9 nm/min. The substrate temperatures given above correspond to the reading of the thermocouple of the substrate heater. Note that the samples B, D, E and F were grown at similar surface temperatures of approximately 800 °C as measured by a calibrated disappearing filament pyrometer. For all samples, first, the substrate was heated to the respective substrate temperature and subsequently, the Ga and the N shutter were opened simultaneously. As already discussed in our previous work [9], prior to the nucleation of the first NWs, a thin TiN layer forms at the surface of the initially sputtered Ti. Here, the formation of the TiN layer was confirmed *in situ* by reflection high-energy electron diffraction (RHEED) and *ex situ* by ellipsometry and energy-dispersive X-ray spectroscopy for all investigated samples (data not shown here).

In order to image the morphology of the NW ensembles presented in this study, micrographs were recorded in a field-emission scanning electron microscope using an acceleration voltage of 5 kV. The length and diameter distributions, as well as the number density of the NW ensembles were determined by analyzing cross-sectional and top-view micrographs with the help of the open-source software ImageJ [48]. Furthermore, the nearest neighbor distances of the NWs in the ensemble were extracted from top-view micrographs as described in Ref [30].

## Results and discussion

The scanning electron (SE) micrographs in Figs. 1 (a)–(c) show the morphology of the NW ensembles A, B, and C grown at different

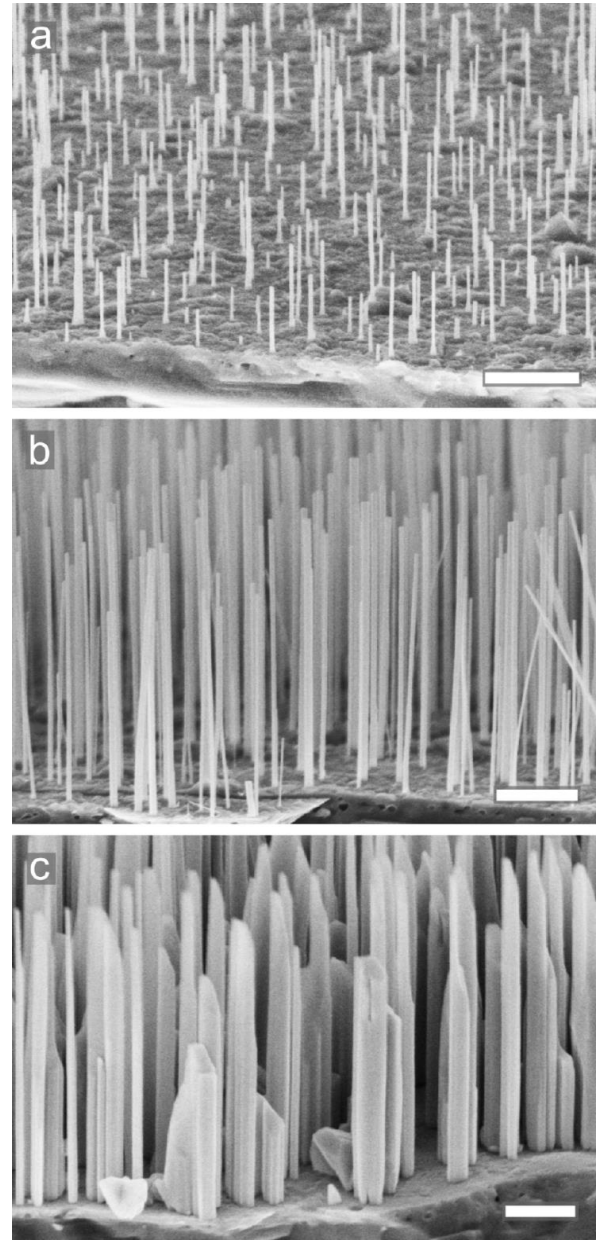

**Figure 1** (a)–(c) Scanning electron micrographs of the NW ensembles A, B, and C grown at three different substrate temperatures 750, 725, and 700 °C at otherwise similar conditions. The NW densities are $1.3 \times 10^9$ cm$^{-2}$, $2.2 \times 10^9$ cm$^{-2}$ and $1.1 \times 10^9$ cm$^{-2}$, respectively. The scale bars correspond to 500nm.

substrate temperatures. The images reveal that changes in $T_{sub}$ of about 25 °C have a strong influence on the final morphology of the NW ensembles on TiN. In the following, we discuss the differences in morphology on the basis of the quantitative analysis of the NW density, diameter and length. We find that with decreasing $T_{sub}$, the number density initially increases from about $1.3\times10^9$ cm$^{-2}$ to $2.2\times10^9$ cm$^{-2}$ and is then reduced to about $1.1\times10^9$ cm$^{-2}$. The lowering of the number density for sample C is attributed to the coalescence of NWs, which leads to the formation of aggregates with larger diameters [7, 25–28, 30]. Consequently, in Fig. 2 (a) we observe in the diameter distribution of sample C with a mean diameter of 67 nm a clear skew with an extended tail towards larger diameters. In contrast, the diameter distributions of sample A and B with mean diameters of 22 nm and 42 nm, respectively, exhibit narrower and more symmetric distributions, signifying the absence of coalescence in these two ensembles [28]. A detailed analysis of the coalescence degree of a NW ensemble on TiN grown under optimized growth conditions will be presented later.

The smaller NW diameters of sample A can be explained by a smaller self-equilibrated NW diameter due to an increased Ga adatom desorption at higher substrate temperatures [49]. More specifically, since during the early growth stage the NWs are able to collect a significant amount of Ga from the surface, the Ga adatom concentration at their top facets exceeds the one of N. As a consequence, radial growth sets in and the effective III/V ratio decreases with increasing diameter until stoichiometry at the top facet is reached. For higher Ga desorption, as expected for the high $T_{sub}$ of sample A, the initial effective III/V flux ratio at the top facet of the NWs is lower and hence stoichiometry is reached earlier, resulting in a smaller self-equilibrated diameter. Indications for increasing Ga desorption are also seen in the length histograms in Fig. 2 (b). The mean length of the NWs decreases with increasing $T_{sub}$, while the width of the length distribution broadens. These trends can be explained as follows. The longer the NWs become during the growth, the smaller is the amount of Ga which is collected from the surface and reaches the NW tips by diffusion along the NW side facets. Hence, once the self-equilibrated diameter and with that stoichiometry at the top facet is reached, due to the elevated substrate temperatures, the effective III/V ratio may become smaller than one. As a result, the axial growth rate may become Ga-limited and decreases with increasing substrate temperature. Furthermore, at high substrate temperatures the increased Ga desorption and GaN decomposition lower the probability of NW nucleation. Hence, the individual NWs nucleate at very different times, resulting in a broader length distribution as it is visible for sample A. The nucleation process of GaN NWs on TiN is discussed in more detail later.

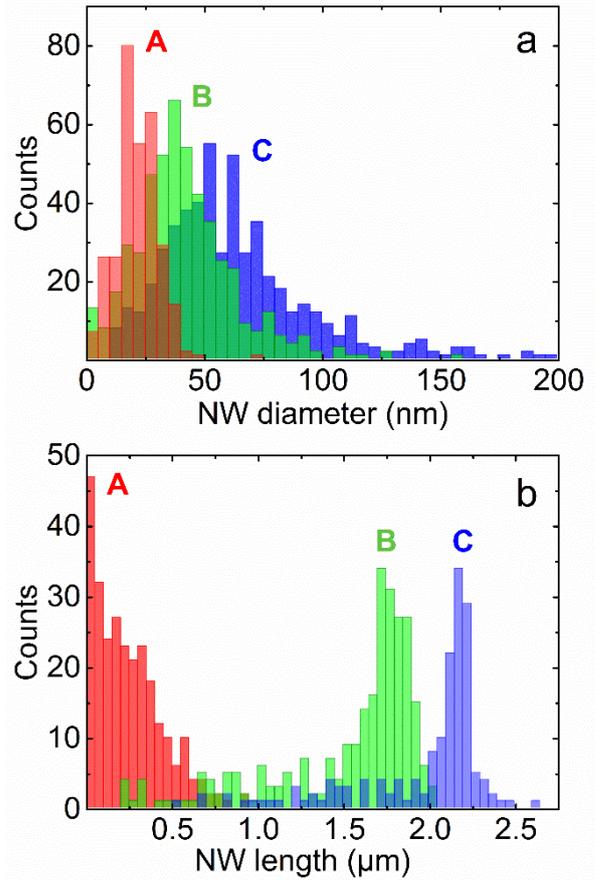

**Figure 2** The histograms in (a) and (b) show the length and diameter distributions of samples A, B, and C with mean diameters of 22, 42 and 67 nm and mean lengths of 220 nm, 1.5 μm and 2.0 μm, respectively.

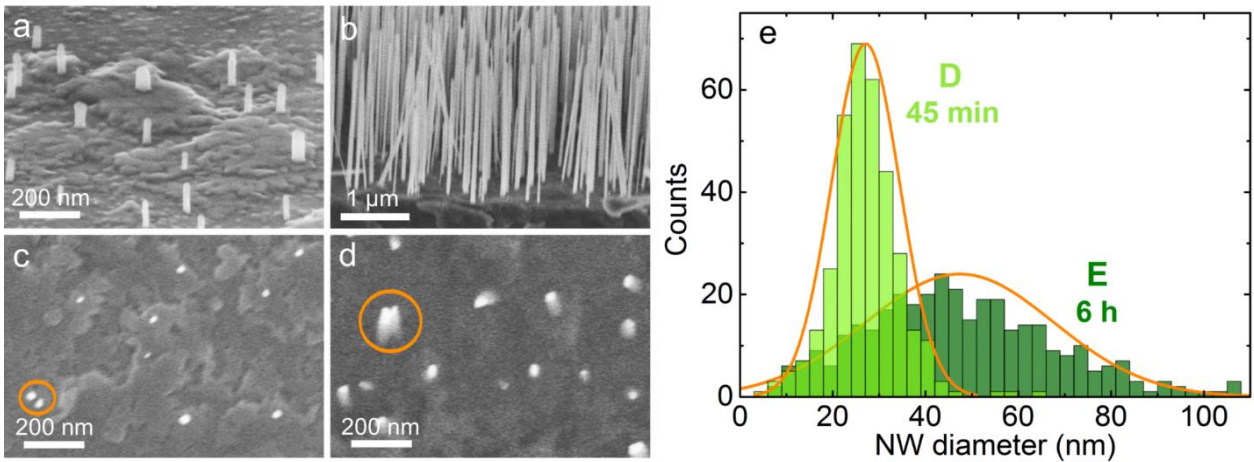

**Figure 3** Bird's eye [(a) and (b)] and top view [(c) and (d)] micrographs of NW ensembles D [(a) and (c)] and E [(b) and (d)] grown at the optimum substrate temperature for 45 min and 6 h, respectively. The corresponding number densities are $5.9\times10^8$ cm$^{-2}$ and $1.1\times10^9$ cm$^{-2}$, respectively. The histogram in (e) shows the diameter distributions of the two NW ensembles. The diameters of the NWs in both samples follow a Gaussian distribution (orange graphs), indicating a low coalescence degree of these NW ensembles.

In general, from the temperature series shown in Fig. 1, it follows that for low substrate temperatures around 700 °C (sample C), one obtains NW ensembles which are rather uniform in length. However, these NWs are fairly thick and coalesced, similar to those grown on Si [28]. For high $T_{sub}$ of around 750 °C (sample A), we found NW ensembles with very thin NWs that do not exhibit any obvious sign of coalescence. At the same time, they are rather short and very inhomogeneous in length, similar to what we previously observed for growth on TiN [9]. In contrast, for the growth at an optimum intermediate $T_{sub}$ (725 °C for sample B corresponding to about 800 °C real surface temperature), NW ensembles with long and thin NWs which are homogeneous in length can be obtained. Most importantly, these NWs appear to be uncoalesced, and we will analyze this aspect in more detail below. Furthermore, very little parasitic growth is seen between these NWs grown on a conductive substrate. So far, the self-assembled formation of GaN NWs combining all these beneficial properties has been very difficult.

In order to elucidate the origin of these beneficial characteristics, we study in the following the evolution of the morphology at the optimum substrate temperature for different growth times. Figure 3 shows bird's eye [(a) and (b)] and top view [(c) and (d)] micrographs of the NW ensembles of sample D [(a) and (c)] and E [(b) and (d)], which were grown for 45 min and 6 h, reaching a mean NW length of about 120 nm and 1.5 μm, respectively. From the analysis of top-view SE micrographs, we extract a NW number density for the samples D and E of $5.9\times10^8$ cm$^{-2}$ and $1.1\times10^9$ cm$^{-2}$, respectively. Consistent with the results for sample B, both samples D and E exhibit mainly well separated NWs, which already suggests a low coalescence degree. Nevertheless, a closer inspection reveals that NWs may nucleate quite close to each other [see orange circle in Fig. 3 (c)], and such NWs might eventually coalesce during growth and form an elongated aggregate [see orange circle in Fig. 3 (d)]. In order to quantify the degree of coalescence of NW ensembles, analyzing the circularity of the NW top facets as well as the dependence of the top facet area on its perimeter proved to be a good measure for NWs on Si [28, 29]. However, on TiN the slight tilt of the NWs makes such an analysis of the circularity of the NWs impractical. Instead, we manually counted visibly coalesced NWs in top-view micrographs and divided this number by the total number of objects. Thus, we were able to estimate the coalescence degree of sample E to be around 7%, which is about a factor ten lower than what has been reported for NW ensembles of the same length grown on Si [28, 29]. For the short growth time of sample D, hardly any indication of coalesced NWs were found, leading to a coalescence degree of less than 1%.

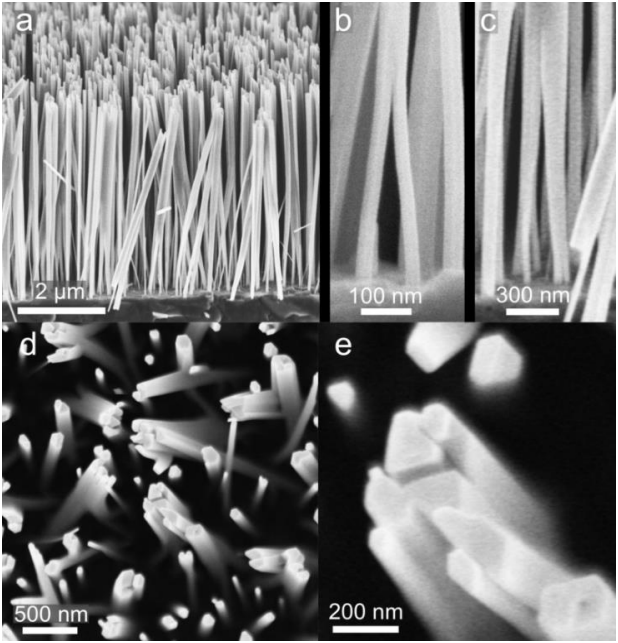

**Figure 4** Bird's eye [(a), (b) and (c)] and top view [(d) and (e)] micrographs of sample F. The mean length of the NWs is with about 4.7 μm three times longer than for sample E. Most of the NWs are coalesced, as the high magnification images (b), (c) and (e) reveal. The sample has been grown for 7 h under the same conditions as the sample presented in Figs. 3 (b) and (d), however, the flux of active N species was twice as high.

Another way to analyze the coalescence of NW ensembles is to consider the distribution of the NW diameters. In particular, the diameter distribution of a NW ensemble with mainly uncoalesced NWs can be well described by a Gaussian distribution [28]. Once the NWs coalesce, they form aggregates of larger diameters. As a result, the diameter distribution of a highly coalesced NW ensemble is clearly skewed with an extended tail towards larger diameter values and is then rather described by a gamma distribution [28, 50]. Figure 3 (e) shows the diameter histograms of samples D and E, with a mean diameter of 24 nm and 43 nm, respectively. The comparatively small diameters of sample D can be explained by the fact that in this early growth stage, most of the NWs have not yet reached their self-equilibriated diameter [49]. As the orange curves indicate, both diameter histograms are fit well by Gaussian distributions. Therefore, the different ways of analysis all show that GaN NW ensembles can be grown on TiN with very low coalescence degree up to a mean NW length of at least 1.5 μm.

Next, we investigate whether coalescence becomes relevant in much longer GaN NWs grown on TiN. To this end, we consider sample F as shown in Figure 4 with a mean NW length of about 4.7 μm. The micrographs already reveal that for NW ensembles of such a length most NWs are coalesced also on TiN. The number density of objects, where one object can be a single NW or an aggregate of several NWs [see Fig. 4 (d)] is about $6.9 \times 10^8$ cm$^{-2}$. Counting the individual wires visible in the thick, merged aggregates, as well as the free-standing NWs in between, the actual number density of nucleated NWs could be estimated to be at least $1.7 \times 10^9$ cm$^{-2}$. Following the same procedure as for samples D and E, a coalescence degree of around 77% was determined. In other words, in NW ensembles grown on TiN with a length of about 5 μm uncoalesced NWs are an exception, similar to the situation on Si for a length of about 1 μm [27–29]. We now analyze the coalescence mechanism in more detail. In a recent study, we showed that on Si the coalescence process is mainly caused by the bundling of NWs and not by the merging of wires due to radial growth and/or their mutual misorientation [30]. Once the NWs reach a certain critical length $H_c$, it is energetically favorable for them to bundle and thus reduce their total surface energy at the expense of the elastic energy of bending. This critical length for coalescence by bundling strongly depends on the NW radii and the distance between the single NWs. The height at which the NWs merge together is then given by $h_c = 3H_c/4$.

Figure 4 reveals that the base of many NW aggregates on TiN consists of at least two initially uncoalesced NWs which bend towards each other and bundle at a certain height, similar to the behavior on Si. By a close inspection of various aggregates, we found that usually the smaller the NW diameter and the distance between the NWs, the smaller is the height at which the NWs coalesce. Figure 4 (b) shows an example where two NWs, which initially nucleated around 100 nm apart from each other, merge at a height of about 425 nm.

Applying the quantitative model of Ref. [30] using diameters of 50 and 60 nm extracted from the SE micrographs for the two NWs, we obtain a height $h_c$ of about 415 nm, i.e., a value very similar to the measured one. As the second example, Figure 4 (c) depicts an event of multiple bundling, demonstrating that bundling also occurs among already coalesced aggregates. Inspecting the base of the two aggregates, we find that they themselves were originally formed by the coalescence of several thin and uncoalesced NWs, which had nucleated very close to each other. Both aggregates with diameters of around 100 nm bundle at a height of about 1.2 μm. Taking into account the distance between the two thick aggregates of 270 nm, the model predicts a value for $h_c$ of around 1.25 μm, which is once again close to the measured value. The two examples in Figure 4 (b) and (c) show that the coalescence of GaN NWs on TiN can be well described by the model of Ref. [30]. We conclude that, in general, the coalescence of GaN NWs on TiN seems to be governed by the same mechanism as on Si.

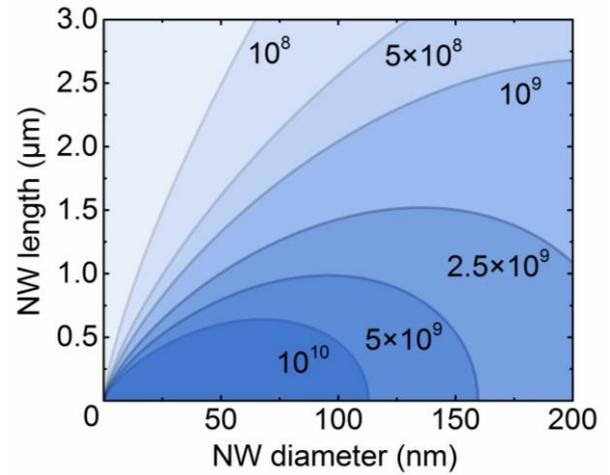

**Figure 5** The different graphs describe the critical length $H_c$ derived in Ref. [30] for NW ensembles with number densities from $10^8$ to $10^{10}$ cm$^{-2}$. Note that here all NWs in the ensemble are assumed to have the same diameter, the same length and to be arranged in a hexagonal lattice at equal distances. Once the NW length in the ensemble exceeds $H_c$, it is energetically more favorable for the NWs to bundle and thus reduce their total surface energy at the expense of the elastic energy of bending. The graphs show that for a given diameter, the NW number density of the ensemble is the crucial factor which enables the growth of well-developed ensembles consisting of long and uncoalesced NWs.

The analysis of samples D, E, and F has shown that for a certain NW number density, the coalescence degree depends mainly on how long the NWs are grown. The graphs in Figure 5 illustrate the critical length $H_c$ in dependence of

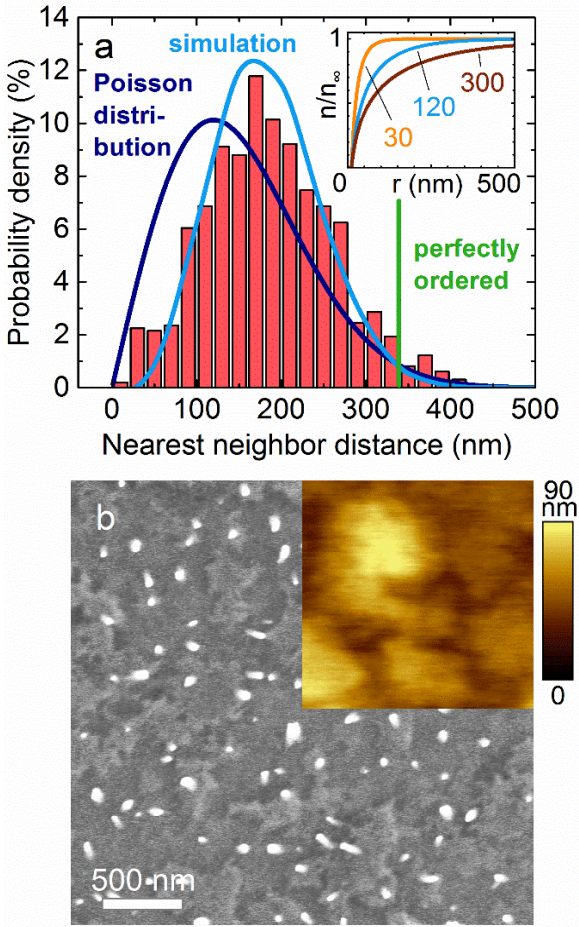

**Figure 6** (a) Analysis of nearest neighbor distances (red histogram) of sample E [see Figs. 3 (b) and (d)]. The vertical green line indicates the NW spacing that would occur in a perfectly ordered NW array with the same number density for a hexagonally closed-packed arrangement. The dark blue curve corresponds to the Poisson distribution that would be found in a NW ensemble with the same number density for completely random and independent nucleation sites. The light blue curve shows the nearest neighbor distance distribution of a simulated NW ensemble for a Ga adatom diffusion length on the surface of $\lambda = 120$ nm and $m = 4$. A detailed explanation of the simulation is given in the text. The inset depicts the behavior of the adatom concentration $n(r)$ around a single NW in dependence on the distance $r$ for different values of $\lambda$: 30, 120, and 300 nm. (b) Top-view SE micrograph of this NW ensemble. The inset depicts an atomic force micrograph of the TiN surface right before NW nucleation at the same magnification.

the NW diameter, above which coalescence sets in for a certain NW number density of the ensemble. For an optimized NW ensemble on TiN with a number density of about $1\times10^9$ cm$^{-2}$ and a mean diameter between 40 and 50 nm, coalescence only occurs for NWs longer than around 1.3 µm. In comparison, for typical NW ensembles on Si, during the early growth stage the NW density rises to more than $1\times10^{10}$ cm$^{-2}$, resulting in coalescence already for very short NWs of only a few hundred nm length [25, 30]. Hence, the key factor for the growth of long and well-developed NW ensembles with low coalescence degree is the low NW number density that can be achieved for the self-assembled formation of GaN NWs on TiN. This consideration raises the question about the origin of the low NW number density on TiN.

In order to answer this question, we now focus on the nucleation of GaN NWs on TiN since it is this process that determines the NW number density. We carried out a statistical analysis of the nearest neighbor (NN) distances between single NWs, and the result for sample E is shown in Figure 6 (a). We observe a broad distribution with values ranging from about 20 to 420 nm with a maximum around 185 nm and a slight tail towards longer NN distances. It is instructive to compare this experimental distribution with the distributions expected for perfect spatial order and complete disorder *for the same NW number density*. If the NWs nucleated in a perfectly ordered array at an equal distance from each other, the NN distance for each NW would be 340 nm (vertical green line). In contrast, if the NWs nucleated independently from each other, the NN distance distribution would follow the Poisson probability density distribution $p(x) = 2\pi\rho x \exp(-\pi\rho x^2)$ (dark blue curve), where $x$ is the NN distance and $\rho$ the number density of the ensemble. Obviously, on TiN neither of these two opposite extremes describes the case of NW ensembles grown on TiN. Nevertheless, the shift of the NN distance distribution towards higher values with respect to the Poisson distribution is a clear indication that NWs do not form completely independently on the surface. Such a non-random nucleation behavior is a natural consequence of a diffusion-induced repulsion of neighboring NWs. In this scenario, once a NW nucleates, the adatoms diffuse towards the nucleus and the adatom concentration in the surrounding area decreases. As a result, the probability for further nucleation close to this NW is reduced.

In the following, we discuss this nucleation process and the role of the Ga adatom diffusion on the surface in more detail. As a starting point, we consider the adatom concentration $n(r)$ at a distance $r$ around an already existing NW. It can be described by the stationary diffusion equation $\lambda^2 \Delta n(r) - n(r) + n_\infty = 0$, where $\lambda$ is the diffusion length of the Ga adatoms on the surface and $n_\infty$ is the adatom concentration far away from the nucleus. Considering the NW to be a perfect sink for Ga adatoms, the boundary condition is $n|_{r=R} = 0$. The solution of this two-dimensional diffusion problem is given by $n(r)/n_\infty = 1 - K_0(r/\lambda)/K_0(R/\lambda)$, where $K_0$ is the modified Bessel function of zeroth order. The inset of Figure 6 (a) shows the behavior of $n(r)/n_\infty$ for three different Ga diffusion lengths of 30, 120, and 300 nm. In the immediate surrounding of the NW, the adatom concentration is small and equal to zero at the interface, but rises continuously with increasing distance $r$ towards one. Moreover, a clear trend can be seen for the influence of the Ga diffusion length $\lambda$; with increasing $\lambda$, the surface area for which the NW affects the nucleation of further NWs is enlarged.

In order to simulate the nucleation behavior of GaN NWs on TiN, we assume that the nucleation probability of a NW directly depends on the Ga adatom concentration on the surface. Using the explicit expression for the adatom concentration derived above, the probability $p(\mathbf{r})$ for a new NW to nucleate at a site $\mathbf{r}$ in between a number of $j$ already existing NWs is assumed to be $p(\mathbf{r}) = \prod_j [n(|\mathbf{r} - \mathbf{r}_j|)/n_\infty]^m$, where $\mathbf{r}_j$ is the position of the $j$th already existing NW and $m$ is a parameter related to the number of Ga atoms contained in the critical nucleus [51–53].

For the simulation of a whole NW ensemble by a Monte Carlo approach, we randomly choose nucleation sites on a predefined area, where NWs nucleate with a respective probability $p(\mathbf{r})$ until a specific NW density is reached. Figure 6 (a) shows that the NN distance distribution of sample E can be very well modeled by the NN distance distribution of a simulated NW ensemble (light blue curve) using a diffusion length of $\lambda = 120$ nm and $m = 4$. The value for $\lambda$ is actually comparable to that reported by Kishino et al. [54], who assessed the diffusion length of Ga adatoms on a nitridized Ti mask for selective-area growth by measuring the distance between single nucleation sites on the mask. The NN distance distribution of a simulated NW ensemble with a smaller adatom diffusion length (not shown here) would be close to the Possion distribution, similar to what has been found for GaN NW ensembles of similar density on Si [30]. For larger adatom diffusion lengths, however, the respective NN distance distribution would be narrower and shifted towards a perfectly ordered arrangement in comparison to sample E.

This analysis shows that the Ga adatom diffusion length on the surface has an important influence on the ordering of the NWs and seems to be mainly responsible for the nucleation behavior on TiN. However, not only the ordering of the NW is influenced, but also the NW number density. The larger the Ga adatom diffusion length, the more unlikely it becomes for a new NW to nucleate close to already existing NWs, which in the end results in a lower NW number density for long and well developed NW ensembles.

Another origin of non-random nucleation, besides diffusion-induced repulsion of neighboring NWs could be heterogeneous nucleation at preferred sites on the surface. Figure 6 (b) shows a top-view SE micrograph that indeed evidences significant roughness of the TiN film on which NW nucleation takes place. The atomic force micrograph displayed in the inset of Figure 6 (b) shows that the surface consists of variously shaped islands of several hundred nm width and length with a peak-to-valley roughness of 90 nm over an area of 1×1 μm. However, a close inspection of several such SE micrographs did not reveal any correlation between the surface morphology and the spatial location of the NWs. The NWs were found to nucleate neither preferentially at the edges, nor on-top nor in between the islands. Hence, diffusion-induced repulsion seems to be the best explanation for the observed NN distance distribution. Interestingly, for the Stranski-Krastanov growth of InAs quantum dots on GaAs Moison et al. [55] found that higher surface roughness shifts the NN distribution towards shorter NN distances and hence closer to the Poisson distribution describing random nucleation. Due to the rather rough surface morphology of the TiN, it is likely that also in our case the diffusion of the Ga adatoms on the substrate and with that the NW density are affected by the surface roughness.

Finally, we test whether the idea of diffusion-induced repulsion of neighboring NWs can explain how the NW density evolves on TiN from the early nucleation stage to the final, fully developed NW ensemble. To this end, we consider again samples D and E grown under the same conditions for different durations (see Fig. 3). For sample D with a growth time of 45 min, most NWs are found to have a similar length of around 120 nm, and there are hardly any short NWs which have just nucleated. This observation suggests that the nucleation density on TiN increases rapidly to a value (here mid $10^8$ cm$^{-2}$) for which most Ga adatoms diffuse to a nearby NW and are incorporated there. As a consequence, the nucleation rate drops significantly. This interpretation is also consistent with the fact that for sample E, which was grown for an additional 5 h, the NW density is only twice as high as for sample D. At the same time, this increase implies that the nucleation of further NWs is not suppressed entirely. Indeed, the histogram of sample B in Fig. 2 (b) reveals the presence of NWs of varying lengths all significantly shorter than the average. One explanation for this ongoing nucleation is that the influence of the already existing NWs on the Ga adatom concentration on the surface decreases once they reach a certain length. Hence, diffusion-induced repulsion and a consequential low (but non-zero) probability of continuous nucleation explain satisfactorily the way in which the NW number density evolves with time on TiN.

We note that this finding is in contrast to the self-assembly of GaN NWs on Si, for which we recently showed that the number density is limited not by adatom diffusion but by shadowing of the substrate from the impinging molecular beams by elongating NWs [30]. On TiN, we estimate that 2 and 70% of the surface are shadowed for sample D and E, respectively. The similar number densities for these samples despite a vast difference in the shadowing efficiency clearly shows that shadowing alone cannot be the main process limiting the number density on TiN. Obviously, shadowing will inevitably contribute to the reduction in nucleation rate with time and completely suppress it for a certain NW length. Still, the key factor that enables number densities as low as $10^9$ cm$^{-2}$ for well developed NW ensembles on TiN is diffusion-induced repulsion.

The comparison of the processes limiting the NW number density on TiN and Si implies that the adatom surface diffusion length is noticeably larger on the former substrate. This result is rather surprising in view of the significant surface roughness of TiN films compared to the atomically smooth Si substrates. Note, however, that the diffusion length of about 120 nm estimated from the modeling of the NN distance distribution in Figure 6 (a) is actually smaller than the typical size of the islands seen in Figure 6 (b). Thus, the main question is rather why the surface diffusion length of Ga adatoms is so short on Si. Further research is needed to develop an atomistic understanding of these diffusion processes.

## Summary and conclusions

In this study, we have demonstrated that during the self-assembled growth of GaN NWs on metallic TiN substrates, the mutual coalescence of NWs is largely absent for NW lengths up to about 1.5 μm. For lengths clearly exceeding this value, coalescence takes place mainly by bundling, similar to what occurs on Si already for much shorter NWs. The key factor leading to coalescence degrees as low as 7% in fully developed NW ensembles on TiN is the low NW number density of about $10^9$ cm$^{-2}$. On TiN, the number density is limited not by shadowing that prevents the nucleation of further NWs on Si, but by the diffusion-induced repulsion of neighboring NWs. In other words, the adatom diffusion length on the substrate is the crucial factor which determines the NW number density and hence the coalescence degree of GaN NWs on TiN.

As a perspective, the properties of the NW ensembles shown in the present work are an excellent basis for the fabrication of more complex structures. Due to the low coalescence degree, the GaN NWs on TiN exhibit a much more regular plan-view cross-section than the typical irregular aggregates on other substrates. This feature may significantly improve the structural quality of axial heterostructures. Finally, let us consider the suitability of self-assembled GaN NW ensembles on TiN for the growth of core-shell heterostructures on the non-polar side facets. High number densities result in strong mutual shadowing of the NWs and thus prevent homogeneous growth on their side facets. In contrast, on TiN the NWs in a fully-developed ensemble with the dimensions of sample E are only shadowed for less than 20% of the time from either Ga or N. Hence, these NW ensembles are promising templates for the realization of well-defined and homogeneous multi-shell heterostructures whose properties may prove useful for applications.

## Acknowledgement

We are grateful to C. Herrmann, H.-P. Schönherr, and C. Stemmler for the maintenance of the MBE system, and A.-K. Bluhm for the SE micrographs. Furthermore, we are thankful to T. Auzelle for a critical reading of the manuscript. Financial support of this work by the Leibniz-Gemeinschaft under Grant SAW-2013-PDI-2 is gratefully acknowledged.
The final publication is available at link.springer.com DOI 10.1007/s12274-017-1717-x.